\documentclass{Interspeech}

\interspeechcameraready

\usepackage{multirow}
\usepackage{makecell}
\usepackage{xcolor}
\usepackage{cite}
\usepackage{microtype}

\newif\ifCommentsAuthors

\usepackage[prependcaption,textsize=scriptsize]{todonotes}
\CommentsAuthorstrue
\ifCommentsAuthors

    \definecolor{myred}{rgb}{.8,.0,.0}

    \definecolor{myblue}{rgb}{0,0.4,.8}

    \definecolor{notecolor}{rgb}{0.4,0.4,.4}

    \definecolor{mcolor}{rgb}{0,0.5,0.1}

    \definecolor{hcolor}{rgb}{0.5,0.2,0.8}

    \definecolor{mygreen}{rgb}{.0,.6,.0}

    \definecolor{hermancolor}{HTML}{FF6600}

    \definecolor{jpcolor}{rgb}{0,0.5,0.5}

    \definecolor{aicolor}{rgb}{.7, 0, 0}
    
\else
    \definecolor{myred}{rgb}{.8,.0,.0}
    \newcommand{\commentmarc}[1]{}
    \newcommand{\commentjulz}[1]{}
    
    \newcommand{\commentbenji}[1]{}
\fi

\title{Analyzing and Improving Speaker Similarity Assessment for Speech Synthesis}

\author[affiliation={1}]{Marc-André}{Carbonneau}
\author[affiliation={1, 2}]{Benjamin}{van Niekerk}
\author[affiliation={1}]{Hugo}{Seuté}
\author[affiliation={1}]{Jean-Philippe}{Letendre}
\author[affiliation={2}]{\\Herman}{Kamper}
\author[affiliation={1}]{Julian}{Zaïdi}

\affiliation{}{Ubisoft La Forge}{Canada}
\affiliation{Electrical and Electronic Engineering}{Stellenbosch University}{South Africa}
\email{marcandre.carbonneau@gmail.com,}
\keywords{speech synthesis, computational paralinguistics}

\usepackage{comment}

\begin{document}

\maketitle

\begin{abstract}
Modeling voice identity is challenging due to its multifaceted nature. 
In generative speech systems, identity is often assessed using automatic speaker verification (ASV) embeddings, designed for discrimination rather than characterizing identity.
This paper investigates which aspects of a voice are captured in such representations.
We find that widely used ASV embeddings
focus mainly on static features like timbre and pitch range, while
neglecting dynamic elements such as rhythm. We also identify
confounding factors that compromise speaker similarity measurements and suggest mitigation strategies.
To address these gaps, we propose U3D, a metric that evaluates speakers’ dynamic rhythm patterns.
This work contributes to the ongoing challenge of assessing
speaker identity consistency in the context of ever-better voice
cloning systems. We publicly release our \href{https://github.com/ubisoft/ubisoft-laforge-spkrid}{code}.

\end{abstract}

\section{Introduction}

Generative models that synthesize virtual humans have progressed tremendously over the last few years.
Virtual characters are expected to possess unique identities that remain consistent across time.
Despite recent progress, this remains a challenge in visual~\cite{gal2022image, ruiz2023dreambooth, huang2024consistentid, wang2024characterfactory} and text-based systems~\cite{lopez-latouche-etal-2023-generating, han-etal-2022-meet}.
This paper considers how identity is measured in generative speech systems.
Voice identity is difficult to define because it spans factors relating to anatomy, such as pitch range and timbre, as well as behavioral factors, affected by a speaker's language, accent, and natural speaking patterns and rhythm~\cite{ladefoged_vowel_1957,johnson2000perception}.
Yet humans routinely identify each other under many changing conditions, implying that a set of measurable markers exists.

Characterizing identity from speech has primarily been studied for automatic speaker verification (ASV) applications \cite{xvect, ECAPA, resnettdnn, wan2018generalized}, where the goal is to discriminate between speakers.
This differs from modeling a speaker's identity~\cite{ulgen2024we}.
For speaker discrimination, the models encode the \textit{minimal} set of markers highlighting the differences between speakers, while a representation useful for generative tasks should encode \textit{all} relevant identity markers.
ASV models can therefore decide to ignore subtle speech identity factors (e.g. pitch patterns, rhythm, and accent) that are crucial for speaker reproduction.

Nevertheless, discriminative ASV representations have been shown to correlate with human judgments \cite{das_predictions_2020}, are successfully used to model voice in few-shot speech synthesis systems \cite{jia2018transfer, cooper_zero-shot_2020, casanova2022yourtts, chien_investigating_2021}, and can be used to align models \cite{hussain2025koeltts}. In addition, these representations are routinely used to measure identity preservation of synthesis systems in research papers \cite{M2Voc2021, chien_investigating_2021, cooper_zero-shot_2020, voxgenesis2024,  wang2024evaluating,  van_niekerk_comparison_2022, hussain2025koeltts, voicebox, VALLE} and community challenges \cite{SVC2023, GenDA2025}. 


In this paper, we examine ASV representations through the lens of synthesis, highlighting their strengths, limitations, and implications for speech generation. 
We begin by reviewing identity markers in speech and identify metrics used to measure them. Next, we analyze how current ASV representations capture these markers. Our findings reveal that ASV embeddings mostly encode spectral information like pitch range and timbre, and that dynamic features are often overlooked.

We then identify confounding factors that can derail speaker similarity experiments.
Specifically, we show that factors such as file duration, channel noise, and equalization can mistakenly cause speech from the same speaker to be attributed to different speakers.
To address this, we provide mitigating strategies and guidelines for experimental protocols and result interpretation.
While we limit our discussion to speech synthesis, our results have broader implications for speaker verification.

To address the shortcoming that ASV embeddings neglect dynamic aspects of identity, we propose U3D (Unit Duration Distribution Distance), a metric to capture speech rhythm patterns.
By modeling rhythm based on phonetic content and as a distribution instead of a statistical moment, we show that U3D captures rich rhythmic information that goes beyond simple metrics like speech rate.
U3D models duration over self-supervised speech units. It, therefore, does not require phonetic annotations, making it language agnostic and practical for many contexts.
We open-source a framework for voice similarity measurement, implementing U3D and our prescribed best practices\footnote{\scriptsize \url{https://github.com/ubisoft/ubisoft-laforge-spkrid}}.



\section{Background}
\subsection{Identity markers in speech}
Foundational phonetic research \cite{ladefoged_vowel_1957, johnson2000perception} identifies two primary categories of cues that enable humans to recognize each other: anatomical and behavioral factors.

Anatomical factors relate primarily to the physical structure of the vocal apparatus. 
The morphology, shape, and size of the vocal tract, along with muscle flexibility, largely define how a person sounds. 
While certain anatomical features remain relatively fixed (such as the dimensions of the nasal cavity), others—notably the oral cavity and pharynx—undergo continuous modification during speech production to modulate phonatory airflow. 
These physical characteristics determine 
the pitch range of a person and their 
timbre. 
Timbre constitutes the distinctive quality that differentiates one voice from another even when lexical content and pitch remain constant~\cite{cleveland_1977}.
Our experiment in Section \ref{sec:exp1} reveals that timbre and pitch are well represented in ASV embeddings, indicating that they play a prominent role in speaker discrimination.  


Behavioral factors encompass learned speech patterns and habits developed by a speaker. 
While also influenced by physical properties \cite{johnson2018}, these factors relate mainly to the manner in which the vocal apparatus is used.
Behavioral factors include 
phoneme inventory and pronunciation habits \cite{resnick1980}, natural rhythm \cite{torgersen2011study, DETERDING2001217}, distinctive phonetic liaisons between words \cite{avanzi2015sociophonetics}, individualized articulation patterns \cite{johnson1993, allen2000}, and speech impediments. 
These behavioral markers reflect influences such as age \cite{foulkes2006}, time-period \cite{thomas2006prosodic}, region \cite{resnick1980, torgersen2011study}, social factors \cite{encreve1983liaison, stuart2014}, and individual traits, all of which play a role in shaping speaker identity.
Despite extensive research on behavioral identity markers, it is still unclear how these should be modeled and measured for speech synthesis applications.
In addition, these complex identity markers are challenging for human evaluators unfamiliar with the speaker \cite{blizzard_2023, PERROTIN2025101747}, limiting the effectiveness of listening tests in assessing the voice similarity of synthesized speech. This underscores the need for automated metrics encompassing  factors beyond anatomical characteristics. 





\subsection{Measuring speaker similarity in synthesis research}
\label{section:protocol}
Recent years have seen growing interest in synthesis evaluation and the scrutiny of experimental protocols \cite{cooper2025goodpractices, kirkland23_ssw, varadhan2024, camp23_interspeech}.
While most efforts focus on naturalness and sound quality, some propose new data and tools for similarity assessments and more~\cite{voxsim24,maimon2025salmon}.

There are different ways to report speaker similarity in synthesis papers. 
Some authors use predictive models to reproduce human annotations \cite{Deja2022, svsnet, maimon2025salmon}.
But most often, speaker similarity is measured by the distance between ASV embeddings, which has been 
shown to correlate with human judgment \cite{das_predictions_2020}. Some authors directly report average similarity between genuine and synthesized speech \cite{VALLE, voicebox, cooper_zero-shot_2020, casanova2022yourtts}. 
However, raw similarity numbers are difficult to interpret because their range varies across ASV representations and they must be put in perspective with similarities between genuine examples of the same speaker. 

This is why many studies report the equal error rate (EER) of an ASV system comparing synthesized utterances with genuine utterances~\cite{das_predictions_2020}.
In practice, reporting EER translates into measuring similarity between pairs of synthesized--genuine utterances and genuine--genuine utterances. 
Then, a decision threshold is set to yield equal rates of false acceptances and false rejections.
If synthetic utterances differ significantly from genuine ones, their similarity scores will be lower, allowing for a decision threshold that gives perfect separation and a 0\% error rate.
Conversely, if synthetic utterances are indistinguishable from genuine ones, it is impossible to set such a threshold, 
making the decision akin to flipping a coin with a 50\% error rate.

\section{Experiments}

While ASV embeddings are widely used for speaker similarity evaluation in synthesis, very little is known about what they actually encode and how robust they are. 
We explore which markers are represented in some widely used ASV embeddings, and measure the effect of confounding factors. Finally, we propose a metric for rhythm that we validate through experimentation. 

\subsection{Models and datasets}

Our goal is not to identify the best ASV technology for speech synthesis applications, but to make observations on a few representative methods. Modern representations rely on neural networks that summarize variable length utterances in fixed-length embedding vectors. The cosine similarity between these vectors indicates the predicted similarity between speakers.

We compare conventional but still widely-used ASV methods such as GE2E~\cite{wan2018generalized} and X-Vectors~\cite{xvect}.
GE2E uses a contrastive loss to learn discriminative embeddings, while X-Vectors are the internal representation of a classifier trained on a finite set of speakers. 
The more recent ResNet-TDNN~\cite{resnettdnn} and ECAPA-TDNN~\cite{ECAPA} models are architectural improvements on X-Vectors, incorporating residual and skip connections and attention for pooling.
We also include more recent methods relying on self-supervised learning for feature extraction because they are increasingly popular and yield state-of-the-art results \cite{huh2024vox}.

We use open-source implementations from SpeechBrain\footnote{\scriptsize \url{https://github.com/speechbrain/speechbrain}}~\cite{speechbrain} for X-Vector, ECAPA-TDNN and ResNet-TDNN, and Resemble-ai's implementation of GE2E\footnote{\scriptsize \url{https://github.com/resemble-ai/Resemblyzer}}.
Finally, we use self-supervised approaches that use WavLM~\cite{wavlm} as feature extractor:
WavLM+ECAPA-TDNN, WavLM Large+ECAPA-TDNN\footnote{\scriptsize \url{https://github.com/microsoft/UniSpeech/tree/main/downstreams/speaker_verification}} and WavLM+X-Vector\footnote{\scriptsize \url{https://huggingface.co/microsoft/wavlm-base-plus-sv}}.

We use the ARCTIC~\cite{kominek04b_ssw} and L2-ARCTIC~\cite{zhao18b_interspeech} speech datasets in all our experiments. ARCTIC comprises recordings of 18 English speakers reading phonetically balanced prompts. L2-ARCTIC expands the number of speakers with 24 non-native English speakers representing a variety of accents. We chose these datasets for their diversity and recording quality. In the experiment of Section \ref{sec:exp1}, we complement our datasets with the 1172 speakers from the LibriSpeech~\cite{Librispeech} train-clean-100 and train-clean-360 subsets, to increase diversity.

\subsection{Markers captured by ASV embeddings}
\label{sec:exp1}

\begin{table*}[ht]
    \centering
    \caption{Extracted speech features and a short explanation and intuition on what they measure.}    
    \scriptsize
    \vspace{-5pt}
    \renewcommand{\arraystretch}{1.2}
    \begin{tabular}{ p{2.6cm} p{6cm} p{7cm} }
        \hline
        \textbf{Markers} & \textbf{Description} & \textbf{Intuition}  \\ \hline 
        
        Duration & Length of the speech example. & This is independent from speaker identity.  \\
        Speech rate & Number of syllables per minute. & Indicates how fast a person speaks. \\
        Voiced segment length & The average duration of a voiced segment. & Indicates how the person elongates vowel-like sounds. \\ 
        Unvoiced segment length & The average duration of an unvoiced segment. & Indicates some pronunciation habits of the speaker.\\ \hline
        
        Mean pitch & Average $F_{0}$ in semitones of voiced segments. & Indicates how high or low a person speaks. \\ 
        Pitch std & Standard deviation of $F_{0}.$ & An expressive speaker will have a high base pitch deviation. \\ \hline
        
        Mean loudness & The average perceived energy in a speech signal. & Mostly indicates of the file normalization level.\\ 
        Loudness std & Standard deviation of perceived energy. & Indicates if speaker is expressive. \\ \hline
        
        Shimmer & Mean amplitude difference between consecutive $F_{0}$ periods. & Voice quality indicator; irregular vocal fold vibrations result in breathiness and can show poor phonation control \cite{wertzner2005analysis}. \\
        Harmonics-to-Noise Ratio (HNR) & Energy ratio between harmonic and noise-like components. & Voice quality indicator; high HNR characterizes sonorant and harmonic voices, while low HNR denotes an asthenic voice and dysphonia~\cite{teixeira2013vocal}.\\
        
        $\alpha$-ratio & Ratio of the energy from 50–1000 Hz and 1–5 kHz. & Reflects voice quality; bright voices have more energy in high frequencies as opposed to dark/mellow voices \cite{Marsano2023}. \\ \hline
        
    \end{tabular}
    \label{tab:feature_description}
\end{table*}

We start by exploring which identity markers are encoded in ASV embeddings.
We do this by trying to predict handcrafted features that capture different aspects of identity from the embeddings. Table \ref{tab:feature_description} summarizes the features and provides an intuition on what they measure. Anatomical characteristics, such as pitch and timbre, are captured by mean pitch and harmonic-to-noise ratio (HNR) and $\alpha$-ratio.
More behavioral and dynamic speech patterns are measured by speech rate, standard deviation of pitch and loudness, as well as voiced and unvoiced segment lengths. We extract speech rate using the method in \cite{van_niekerk_rhythm_2023} because it highly correlates with the ground truth syllable rate. Using REAPER\footnote{\scriptsize \url{https://github.com/google/REAPER}}, we compute the mean and standard deviation (std) of pitch only on voiced segments. The other features are extracted using OpenSmile \cite{opensmile}. Lastly, we include file duration, which should not be linked to speaker identity. 

We extract ASV embeddings for each utterance in LibriSpeech, ARCTIC, and L2-ARCTIC.
Taking the embeddings as input, we train a lasso regressor to predict the features for the utterance.
We also experimented with a non-linear random forest regressor, which yielded results comparable to those of lasso; therefore, we report only the lasso results. We remove outliers using the inter-quartile range (IQR) method and optimized the hyper-parameters by performing 5-fold cross-validation, making sure that every speaker belongs only to a single fold. After retraining the regressor on all data, we predict the value of each handcrafted feature. Finally, we compute the coefficient of determination ($r^2$) between the feature values and their predictions.
A perfectly predicted feature yields a score of 1.0 while uncorrelated predictions yield a score of 0.0. 

\begin{figure}[!t]
    \centering
    \includegraphics[scale=0.5]{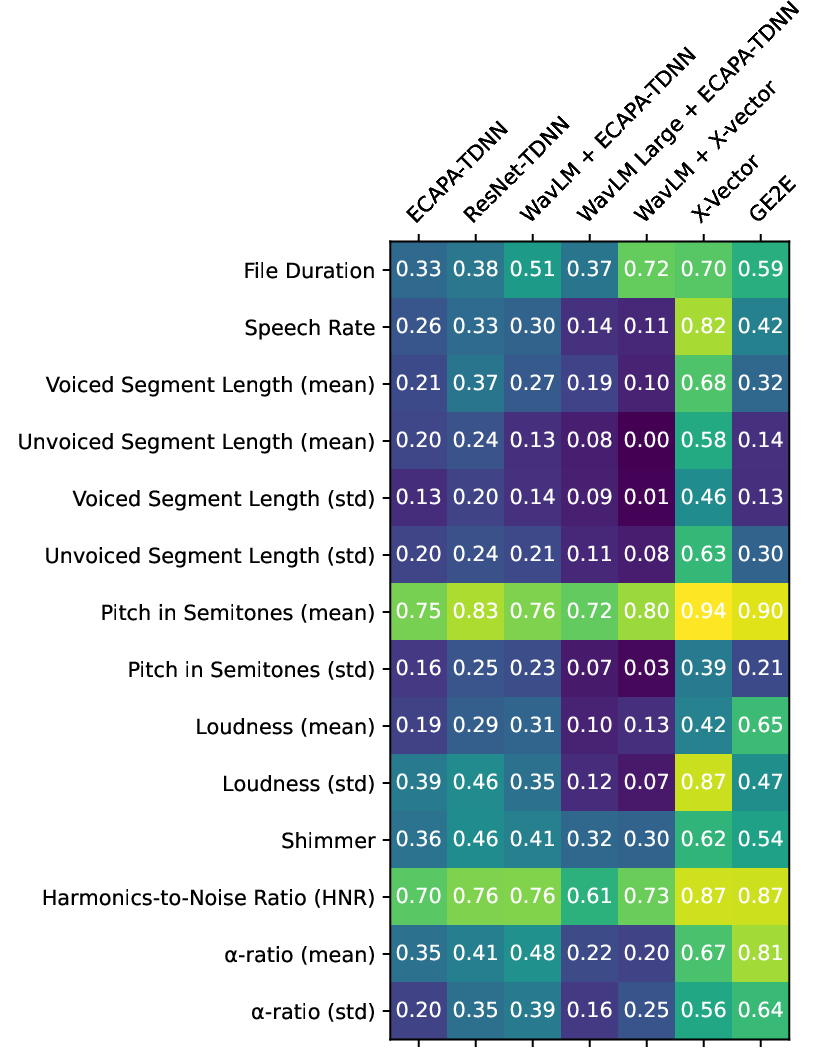}
    \caption{The coefficient of determination ($r^2$) of feature predictions. A perfect score is 1.0 and uncorrelated predictions yield a score of 0.0.}
    \label{fig:colored_table}
\end{figure}


Figure \ref{fig:colored_table} shows that the speaker embeddings primarily encode static spectral information (mean pitch, HNR, shimmer, $\alpha$-ratio) reflecting voice quality and frequency range.
On the other hand, the embeddings fail to capture dynamic behavioral identity markers like speech rate, voiced/unvoiced segment durations, variations in pitch, and loudness. This limitation means that embeddings represent only a 
subset of the speaker characteristics necessary for a comprehensive assessment of identity in speech synthesis.

While our goal is not to identify the best speaker embedding to use for identity measurement, we find that X-Vector embeddings encode a wider range of markers, including more dynamic information, than more recent and better-performing embeddings for the speaker verification task.
This agrees with the findings of \cite{fu2024asrrl}. This somewhat surprising result highlights the differences between speaker verification and speech synthesis, and calls for a more extensive comparison of ASV embeddings in the synthesis context.
Better-suited identity representations might even need to be developed.

Our experiment also shows that embeddings can encode distracting factors unrelated to identity, such as speech file duration. We now explore this limitation and its implications in more detail.

\subsection{Robustness of automatic speaker similarity assessment}

\begin{table*}[ht]
\caption{Mean and standard deviation of EER across ARCTIC and L2-ARCTIC speakers for different experiments.}
    \vspace{-5pt}
    \renewcommand{\arraystretch}{1.2}

\scriptsize
\centering
\begin{tabular}{l c c c c c c c}  


\toprule
\textbf{Experiment}
& ECAPA-TDNN 
& ResNet-TDNN 
& \makecell[c]{WavLM \\ +ECAPA-TDNN} 
& \makecell[c]{WavLM Large \\ +ECAPA-TDNN} 
& \makecell[c]{WavLM \\ +X-Vector} 
& X-Vector 
& GE2E \\
\hline  
One speaker \textit{vs} rest  & $0.00 \pm 0.01$  & $0.00 \pm 0.00$  & $0.00 \pm 0.00$  & $0.00 \pm 0.00$  & $0.05 \pm 0.03$  & $0.03 \pm 0.01$  & $0.02 \pm 0.01$ \\
Same speaker & $0.50 \pm 0.03$  & $0.50 \pm 0.02$  & $0.51 \pm 0.02$  & $0.50 \pm 0.02$  & $0.50 \pm 0.03$  & $0.50 \pm 0.03$  & $0.50 \pm 0.03$ \\
Same speaker, short \textit{vs} long & $0.34 \pm 0.04$  & $0.35 \pm 0.03$  & $0.32 \pm 0.03$  & $0.33 \pm 0.03$  & $0.39 \pm 0.02$  & $0.30 \pm 0.03$  & $0.34 \pm 0.03$ \\
\hline
SNR 40 & $0.47 \pm 0.03$ & $0.47 \pm 0.03$ & $0.47 \pm 0.03$ & $0.48 \pm 0.03$ & $0.46 \pm 0.04$ & $0.45 \pm 0.05$ & $0.42 \pm 0.04$ \\
SNR 20& $0.34 \pm 0.06$ & $0.32 \pm 0.05$ & $0.33 \pm 0.06$ & $0.38 \pm 0.05$ & $0.36 \pm 0.07$ & $0.21 \pm 0.07$ & $0.15 \pm 0.06$ \\
SNR 0 & $0.05 \pm 0.04$ & $0.04 \pm 0.04$ & $0.05 \pm 0.03$ & $0.10 \pm 0.05$ & $0.06 \pm 0.04$ & $0.01 \pm 0.01$ & $0.01 \pm 0.01$ \\
\hline
+emphasis & $0.47 \pm 0.04$ & $0.48 \pm 0.03$ & $0.45 \pm 0.02$ & $0.43 \pm 0.03$ & $0.48 \pm 0.03$ & $0.47 \pm 0.04$ & $0.07 \pm 0.03$  \\
+de-emphasis & $0.38 \pm 0.05$ & $0.41 \pm 0.05$ & $0.37 \pm 0.07$ & $0.32 \pm 0.06$ & $0.44 \pm 0.06$ & $0.28 \pm 0.07$ & $0.01 \pm 0.01$ \\
+emphasis +re-equalization & $0.50 \pm 0.02$ & $0.49 \pm 0.02$ & $0.49 \pm 0.02$ & $0.44 \pm 0.04$ & $0.50 \pm 0.03$ & $0.50 \pm 0.03$ & $0.50 \pm 0.02$ \\
+de-emphasis +re-equalization & $0.50 \pm 0.03$ & $0.50 \pm 0.02$ & $0.49 \pm 0.03$ & $0.45 \pm 0.05$ & $0.50 \pm 0.02$ & $0.49 \pm 0.02$ & $0.50 \pm 0.02$ \\
\bottomrule
\end{tabular}
\label{table:eer_table}
\end{table*}

In this section we demonstrate how ASV-based comparisons can be compromised by confounding variables, particularly file duration and channel characteristics.
We do not want to invalidate established methodologies, but rather aim to guide researchers in avoiding potential methodological errors in interpreting experimental results.

Here we measure the EER (see Section \ref{section:protocol}) on the ARCTIC and L2-ARCTIC datasets under various conditions.
Table \ref{table:eer_table} reports the average EER with the standard deviation over all speakers.
As reference, we report EER after comparing each speaker to the other speakers, akin to a speaker verification task. A perfect ASV embedding would obtain an EER of 0\%. As a second reference experiment, we consider each speaker individually against themselves: For each speaker, we randomly distribute speech files between two groups, the first acting as \textit{genuine} and the second acting as \textit{synthesized} examples. Being from the same person, both groups should be indistinguishable and EER should be 50\%. Next, instead of randomly distributing the utterances, we sort them by duration, grouping shorter utterances together in one group and longer ones in another.
We then measure EER for discriminating between shorter and longer utterances.
Since these are the same utterances spoken by the same speaker, EER should be 50\%.

As expected, control experiments with same speaker utterances yield an average EER of around 50\% for all methods, and EERs between 0\% and 5\% when comparing different persons. This validates our protocol and shows that all embeddings accurately discriminate speakers.
However, when using the same utterances but sorted by duration, EER is significantly lower for all methods. This means that these ASV systems can discriminate utterances based on file duration, which is not an indicator of identity. This coincides with the results in Figure~\ref{fig:colored_table}, showing that ASV embeddings encode duration. This undesirable behavior can derail experiments if the duration of the synthesized examples is very different from the genuine examples:
a synthesis system would under-perform, and comparisons between systems might not be meaningful.
Fortunately, this can be easily mitigated with careful experimental design that ensures similar utterance durations. In the case of text-to-speech evaluation, we recommend generating the same text utterances as the genuine corpus, taking the usual precaution of not using utterances used during training. 

We next investigate channel-related confounding factors, specifically noise and equalization effects. 
This is important for synthesis evaluation as sound quality varies across methods. While clean, noise-free speech is desirable, quality differences should not influence speaker similarity measurements.
Some systems produce artifacts that reduce sound quality compared to authentic speech, while others generate samples that are cleaner than the enrollment recordings. This latter scenario frequently occurs in voice cloning applications where users record enrollment samples with consumer-grade equipment. The 2023 Blizzard Challenge demonstrated this phenomenon, with several competitors producing samples superior to the training material \cite{blizzard_2023}. 
To assess noise effects, we progressively introduce white noise to speech samples from each speaker and calculate EER when comparing identical utterances at varying signal-to-noise ratios (SNR). The middle section of Table \ref{table:eer_table} presents the average EER across speakers.

Similarly, we measure the effect of equalizing files by applying a popular emphasis and de-emphasis filter defined by $H(z)=1-\alpha z^{-1}$ with $\alpha=0.97$ for emphasis, or its inverse for de-emphasis. These filters modify the spectral balance of the signal, without altering the identity perceived by humans. Results are given in the bottom section of Table \ref{table:eer_table}.

Our experiments on channel noise (Table \ref{table:eer_table} middle section) 
reveal significant variations in EER following perturbation. 
At low SNR, all models incorrectly classify noisy samples as different identities from their clean counterparts. 
While 0 dB SNR represents extreme degradation, the results demonstrate that noise
significantly impacts similarity scores measured by ASV systems.
This has important implications when comparing systems with differing output quality, for instance when evaluating high-performance voice cloning systems using data collected in uncontrolled environments.
\textbf{We recommend that researchers conduct SNR assessments on both genuine and synthesized speech} to ensure valid interpretation of the results. 

Altering the spectral balance of the recording by applying equalization (Table \ref{table:eer_table} bottom section) significantly alters EER for most methods.
For the GE2E implementation that we use, this even leads to catastrophic failure.
Practical synthesis systems, for various reasons, can color the speech samples, making evaluation problematic. \textbf{We suggest to re-equalize the samples from the evaluated system} to get a similar spectral balance as the genuine samples. 
As an example, we mitigate tonal imbalance in our experiment by using a simple equalization matching algorithm. We start by estimating the power spectral densities of the genuine audio samples $S_{rr}(f)$ as well as the emphasized/de-emphasized audio samples $S_{xx}(f)$. We compute the frequency response of the corrective equalization as $|G(f)|^2 = S_{rr}(f)/S_{xx}(f)$. Then, we obtain a 16-band FIR graphic equalizer filter \cite{valimaki2016all}, matching this frequency response. This filter is applied to the emphasized/de-emphasized samples to re-equalize, ensuring that it adopts the tonal characteristics of the genuine examples. In the last two rows of Table \ref{table:eer_table}, we see that this brings EER back to the expected 50\%.



\subsection{Measuring rhythm}

\begin{figure}[!b]
    \centering
    \includegraphics[width=\columnwidth]{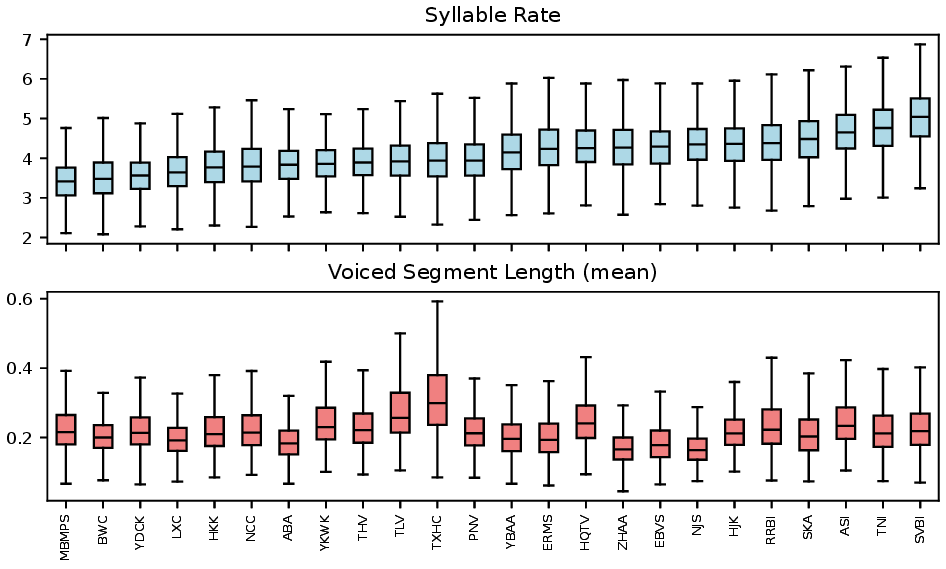}
    \caption{Ground-truth syllable rate and voiced segment length for all speaker in L2-ARCTIC.
    The syllable rates are similar between many speakers, and voiced segment length (bottom) is not strongly correlated with speech rate (top).}
    \label{fig:boxplots}
\end{figure}

We have shown in Section~\ref{sec:exp1} that ASV-based similarity evaluations only capture a subset of identity markers. They are particularly poor at measuring dynamic features. As a first step towards a better characterization of behavioral identity, we introduce U3D (Unit Duration Distribution Distance), a novel metric for modeling rhythm and phoneme duration.
First, we argue that a coarse approach to measuring rhythm, based on speaking rate or mean voiced/unvoiced segment length, is insufficient for differentiating dynamic speech patterns across speakers
To illustrate this, Figure \ref{fig:boxplots} shows the distributions of ground-truth syllable rate and mean voiced/unvoiced segment length for the L2-ARCTIC dataset (which includes speakers from diverse origins). 
In many cases, adjacent speaker pairs do not show statistically significant differences in syllable rate (top). Moreover, speakers with similar syllable rates can exhibit markedly different lengths of voiced segments (bottom), suggesting that natural rhythm involves more complexity than speech rate alone can capture.
These findings underscore the need for a more sophisticated evaluation method that accounts for content and is more descriptive than first or second statistical moments.

We propose comparing the duration distributions of different phoneme types and silence to capture finer-grained aspects of rhythm.
While phoneme-duration analysis requires a forced aligner--a tool not universally available across languages--our U3D metric offers a more accessible alternative. 
By computing duration distributions over unsupervised speech unit groups (obtained from an SSL model), U3D provides a refined characterization of rhythm, modeling duration variations across different sound and phoneme types. This method offers researchers a robust, language-agnostic technique for quantifying the subtle variations in speakers' temporal speech patterns.


U3D is computed in three steps:
    
\noindent \textbf{Discover phoneme-like groups of speech units:} Following~\cite{van_niekerk_rhythm_2023}, we use agglomerative hierarchical clustering over speech units \cite{van_niekerk_comparison_2022} extracted from a HuBERT model
\footnote{\scriptsize \url{https://github.com/bshall/urhythmic}}. 
Different branches in the resulting dendrogram correspond with broad phoneme-like groups.
For example, \cite{van_niekerk_rhythm_2023} show that the three main clusters map to sonorants, obstruents, and silences. Refer to their paper for a visualization.

\noindent \textbf{Segment speech and compute duration distributions:} First, we extract speech units from all speech utterances.
Then, we partition them into sequences of contiguous segments using the score function from \cite{van_niekerk_rhythm_2023}. 
Finally, we record the segment durations to get a distribution for each phoneme-like group.

\noindent \textbf{Measure distance between distributions:} 
For each group, we compute the Wasserstein distance between the duration distributions of the synthesized and genuine speech.
These distances can be reported separately for fine-grained analysis, or averaged to a give a single number for comparison.

To validate our method, we first measure duration distributions using ground-truth phoneme and silence segments obtained through forced alignment\footnote{\scriptsize\url{https://github.com/MontrealCorpusTools/Montreal-Forced-Aligner}}.
We categorize phonemes into five groups: vowels, approximants, nasals, fricatives, and stops. 
Then, we compute average distances across ARCTIC and L2-ARCTIC speakers under three scenarios:
\begin{enumerate} 
    \item \textbf{Same:} Compare distances between two randomly split subsets of a single speaker.
    This reflects the lower bound of our metric, where we expect the smallest distances.
    \item \textbf{Nearest:} Calculate distance between each speaker and their closest counterpart by syllable rate. This setting tests whether our metric can distinguish between speakers undistinguishable with speech rate.
    \item \textbf{Random:} Report distance between random speaker pairs.
\end{enumerate}

\begin{table}[h]
    
    \centering
    \caption{Average Wasserstein distance based on forced alignments and unsupervised speech units (U3D) for the \textbf{same} speaker, between a speaker and their \textbf{nearest} neighbor by syllable rate, and between \textbf{random} pairs of speakers.}
    \vspace{-5pt}
    \renewcommand{\arraystretch}{1.2}
    
    \label{tab:dist_FA}
    \scriptsize
    \begin{tabular}{lccccccc}
        \toprule
        & approx. & fric. & nasal & stop & vowel & sil. & Avg. \\
        \midrule
        \multicolumn{7}{l}{\textbf{Forced Aligned}} \\
        Same & 1.7 & 1.5 & 1.7 & 1.3 & 1.4 & 7.3 & 2.48  \\
        Nearest & 8.6 & 8.5 & 7.4 & 7.4 & 7.6 & 70.7 & 18.37 \\
        Random & 11.7 & 13.9 & 12.9 & 11.5 & 15.0 & 81.6 & 24.43 \\

        \midrule
        \multicolumn{7}{l}{\textbf{Unsupervised (U3D)}} \\
        Same & 2.3 & 1.0 & 1.5 & 1.2 & 1.0 & 5.9 & 2.15 \\
        Nearest & 11.2 & 4.2 & 5.9 & 12.0 & 5.2 & 71.9 & 18.40 \\
        Random & 15.4 & 7.0 & 10.7 & 15.6 & 9.1 & 71.4 & 21.53 \\
        \bottomrule
    \end{tabular}
\end{table}

The results in the top section of Table \ref{tab:dist_FA} show that rhythm distances are significantly larger between different speakers, even those with similar speech rates.
Our approach therefore captures meaningful differences between speakers that cannot be reduced to speech rate.
We subsequently (bottom section) replicate the experiment using unsupervised speech unit groups, confirming our method does not have to rely on forced alignment. This provides a flexible, language-agnostic technique for quantifying subtle variations in speakers' temporal speech patterns.

\section{Conclusion}
In this paper, we explored the limitations of ASV embeddings in the context of assessing speaker identity in speech synthesis.
We showed that ASV embeddings mainly encode speech identity markers relating to anatomy (e.g. pitch range and timbre), and fail to capture time-dependent behavioral identity cues. Next, we explored the robustness of the embedding to distraction factors such as sample duration, noise level, and equalization. We provide recommendations for experimental protocol design and meaningful result interpretation.
Finally, to address the shortcoming of ASV-based comparisons, we proposed U3D, a metric that better characterizes dynamic aspects of a identity by measuring rhythm.
Future work includes additional metrics for behavioral markers, an extensive comparison of speaker embeddings, novel representations designed specifically for synthesis evaluation, and an extension of the discussion beyond neutral speech.
\bibliographystyle{IEEEtran}

\bibliography{paper}

\end{document}